\documentclass[fleqn,twoside]{article}
\usepackage{espcrc2}
\usepackage{epsfig}

\usepackage{graphicx}
\usepackage[figuresright]{rotating}

\newcommand{\AmS}{{\protect\the\textfont2
  A\kern-.1667em\lower.5ex\hbox{M}\kern-.125emS}}

\title{OPE Constraints and the Leading Order Hadronic Contribution to
$(g-2)_\mu$}

\author{K. Maltman\address{Department of Math and Stats, York Univ., 
Toronto, ON, M3J 1P3, Canada, and
CSSM, Univ. of Adelaide, Adelaide, SA, 5005, Australia}
\thanks{e-mail: kmaltman@yorku.ca. Work supported by a grant from
the Natural Sciences and Engineering Research Council of Canada}}
       
\begin{document}

\begin{abstract}
OPE constraints are studied as a means of distinguishing 
between the versions of the $I=1$ vector spectral function extracted from
(i) inclusive $I=1$ hadronic electroproduction cross-sections
and (ii) inclusive $I=1$ hadronic $\tau$ decay data, with the goal
of clarifying expectations for the leading order hadronic
contribution to $a_\mu = (g-2)_\mu /2$ in the Standard Model. 
The constraints are shown to, at present,
favor the $\tau$ decay data, and hence a Standard Model 
prediction for $a_\mu$ compatible with the BNL E821 experimental result. 
The relative role of the $\pi\pi$ and $4\pi$ contributions
to the discrepancy between the integrated electroproduction 
results and the corresponding OPE constraints is also investigated,
and the significance, in this context, of forthcoming re-measurements
of the $e^+e^-\rightarrow \pi^+\pi^-\pi^0\pi^0$ cross-sections
pointed out.
\vspace{1pc}
\end{abstract}

\maketitle

\section{Background}
It is well known that
\begin{itemize}
\item the leading order (LO) hadronic contribution
to $a_\mu$, $\left[ a_\mu \right]^{had,LO}$, can be written as a
dispersive integral
\begin{equation}
\left[ a_\mu\right]^{had,LO}\, =\, {\frac{\alpha_{EM}^2}{3\pi^2}}\,
\int_{4m_\pi^2}^\infty ds\, {\frac{K(s)}{s}}\, R(s)
\label{amudispersive}\end{equation}
with $K(s)$ a known function and 
\begin{equation}
R(s)\, =\, {\frac{3s\, \sigma [e^+e^-\rightarrow hadrons]}{16\pi\, 
\alpha_{EM}^2}}
\end{equation}
\item as a consequence of CVC, 
the $I=1$ parts of both $R(s)$ and the hadronic $\tau$ decay 
distribution provide experimental determinations of
the $I=1$ vector spectral function, $\rho^{I=1}(s)$, allowing $\tau$
decay data to, in principle, be incorporated into the determination of
$\left[ a_\mu \right]^{had,LO}$~\cite{davierusetau}
\item after known isospin-breaking (IB) corrections to the
CVC relation have been performed, the current $\tau$ and electroproduction (EM)
versions of $\rho^{I=1}(s)$ do not agree~\cite{dehz}, leading to
incompatible determinations of $\left[ a_\mu \right]^{had,LO}$~\cite{dehz06}.
\end{itemize}
The $\tau$-based determination corresponds to a SM value
for $a_\mu$ compatible, within errors,
with the BNL E821 experimental result~\cite{bnlgminus2}, 
while the most recent assessment of the electroproduction (EM) determination
leads to a value in $\sim 3\sigma$ disagreement
with experiment~\cite{dehz06}.

The discrepancy between the measured EM version of 
$\rho^{I=1}(s)$ and that implied by IB-corrected $\tau$ decay data is
now generated essentially entirely by contributions from 
the $\pi^+\pi^-$ and $\pi^+\pi^-\pi^0\pi^0$ states. The differences,
moreover, are such that, where the two $\rho^{I=1}(s)$ determinations
disagree, that based on the $\tau$ data lies higher.
Figures of the preliminary BELLE $\tau^-\rightarrow \pi^-\pi^0\nu_\tau$ 
data~\cite{belletaupipi} indicate that, while the region over which a 
discrepancy is observed to occur may be shifted by the BELLE results, it 
remains true that, where disagreement exists, it is the $\tau$ version 
which is higher. 

As pointed out in Ref.~\cite{kmamu06}, the uniformity in sign of
$\rho^{I=1}_{\tau}(s) -\rho^{I=1}_{EM}(s)$ makes it rather easy to construct 
finite energy sum rules (FESR's) which distinguish
between the two versions of $\rho^{I=1}(s)$. The form of these FESR's is
\begin{equation}
\int_0^{s_0}w(s) \rho(s) ds \, =\, - {\frac{1}{2\pi}}\oint_{\vert
s\vert =s_0}w(s) \Pi (s) ds
\label{basicfesr}\end{equation}
where $w(s)$ in an analytic weight function, $\rho (s)\equiv \rho^{I=1}(s)$,
and $\Pi (s)$ is the relevent (charged or neutral current)
$I=1$ vector current correlator. The parameter $s_0$ 
is to be chosen large enough that the OPE representation of $\Pi$ 
can be used on the RHS of Eq.~(\ref{basicfesr}). For weights $w(y)$ 
which are (i) functions of the dimensionless variable $y=s/s_0$ 
and (ii) both non-negative and monotonically decreasing on the 
interval $0\leq y\leq 1$, it follows that, if the $\tau$ data 
is correct, then both the slope with respect to $s_0$ and the magnitude 
of the EM spectral integrals {\it for all $s_0$} will be too small,
relative to OPE expectations. Similarly, if the EM data is correct, 
then both the magnitude and slope with respect to 
$s_0$ of the $\tau$ spectral integrals will be too large, relative to
OPE expectations. It is worth stressing that, for the OPE integrals,
the slope with respect to $s_0$ is considerably less sensitive 
than is the normalization to the main OPE input, $\alpha_s$. 
We will return to this point in the discussion below.

In the analysis reported here we restrict our attention to 
$s_0>2\ {\rm GeV}^2$, which choice strongly suppresses residual OPE 
breakdown for weights, $w(s)$, having a zero at $s=s_0$~\cite{kmfesr,cdgm}.
At these scales, the OPE representation is strongly dominated
by the $D=0$ perturbative contribution, and hence well-determined once
$\alpha_s$ is given at some particular reference scale. The needed input
value is obtained by averaging the independent high-scale determinations of
$\alpha_s(M_Z)$ reported in Ref.~\cite{pdg06qcdreview}, which yields
$\alpha_s(M_Z)=0.1198\pm 0.0020$. Values relevant to the lower scales 
required in this analysis are then obtained via 4-loop running and 
matching~\cite{cks97}. Details of the relevant input for, and treatment of, 
higher dimension OPE contributions may be found in Ref.~\cite{kmamu06}.
The weights used below, $\hat{w}(y)=1-y$ and 
$w_N(y)=1-\left({\frac{N}{N-1}}\right) y+\left({\frac{1}{N-1}}\right) y^N$,
$N=3,\cdots ,6$, are chosen to strongly reduce sensitivity to potential 
poorly known $D\geq 6$ OPE contributions~\cite{kmamu06}. 
The $w_N(y)$ also have a double zero at $s=s_0$, further suppressing 
possible residual OPE breaking contributions. 

The $I=1$ hadronic $\tau$ decay distribution has been measured by 
ALEPH~\cite{alephud}, CLEO~\cite{cleoud} and OPAL~\cite{opalud}. 
The results below are based on the ALEPH determination, 
for which the covariance matrix has been made
publicly accessible. Additional corrections for long-distance EM effects, as
evaluated in Refs.~\cite{cen}, are applied to the $\pi\pi$ component
of the $\tau$ data~\cite{thanksvc}. For pre-2003 EM data, we employ the results
for various exclusive modes reported in the compilation of 
Ref.~\cite{whalley03}. Small missing modes are accounted for
using isospin relations and the techniques described in detail
for such modes in Ref.~\cite{dehz}. Post-2003 updates for a number of 
modes~\cite{newerem} have also been taken into account.
Where uncertainties exist in older publications concerning the treatment 
of radiative corrections and/or systematic errors, and newer data with no such
uncertainties are available, we employ only the newer data.

\section{Results}
\begin{figure}
\unitlength1cm
\caption{EM OPE and spectral integrals for $\hat{w}(y)$} 
\begin{minipage}[t]{8.0cm}
\begin{picture}(7.9,7.9)
\epsfig{figure=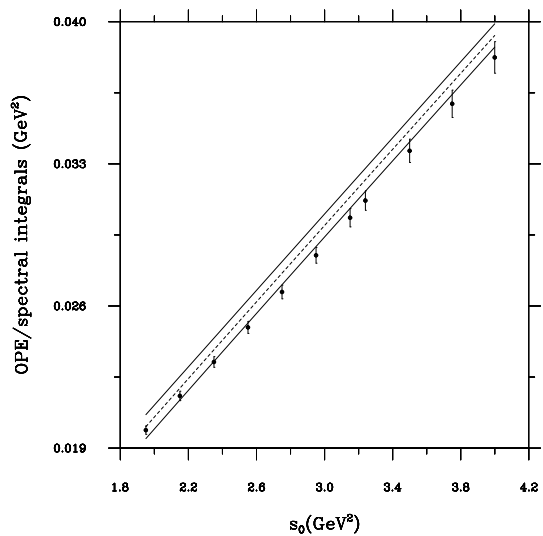,height=7.8cm,width=7.8cm}
\end{picture}
\end{minipage}
\label{fig1em}\end{figure}

\begin{figure}
\unitlength1cm
\caption{EM OPE and spectral integrals for $w_6(y)$}
\begin{minipage}[t]{8.0cm}
\begin{picture}(7.9,7.9)
\epsfig{figure=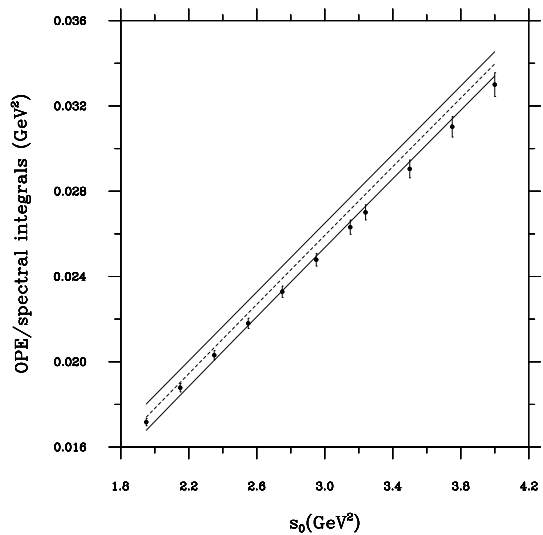,height=7.8cm,width=7.8cm}
\end{picture}
\end{minipage}
\label{fig2em}\end{figure}

\begin{figure}
\unitlength1cm
\caption{$\tau$ OPE and spectral integrals for $\hat{w}(y)$}
\begin{minipage}[t]{8.0cm}
\begin{picture}(7.9,7.9)
\epsfig{figure=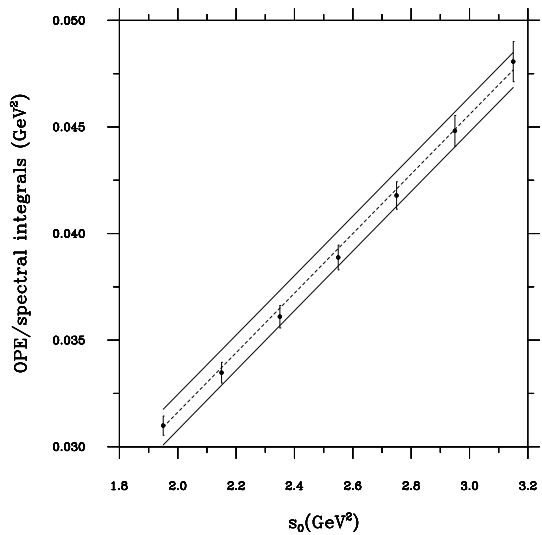,height=7.8cm,width=7.8cm}
\end{picture}
\end{minipage}
\label{fig1tau}\end{figure}

\begin{figure}
\unitlength1cm
\caption{$\tau$ OPE and spectral integrals for $w_6(y)$}
\begin{minipage}[t]{8.0cm}
\begin{picture}(7.9,7.9)
\epsfig{figure=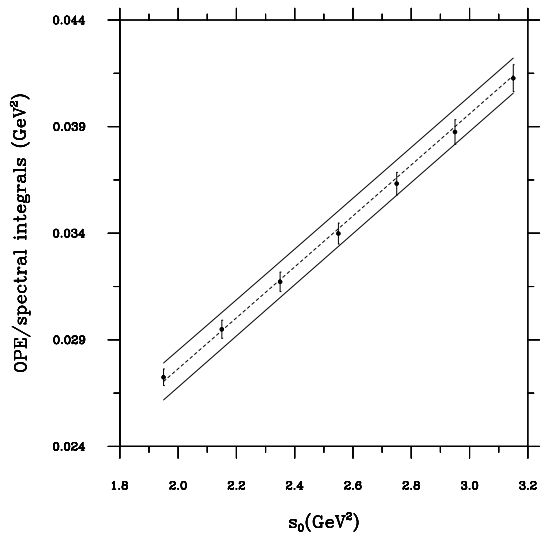,height=7.8cm,width=7.8cm}
\end{picture}
\end{minipage}
\label{fig2tau}\end{figure}

Examples of the results of the comparison between the spectral integrals
and OPE expectations, for the EM case, are shown in Figures \ref{fig1em} and 
\ref{fig2em}, for the weights $\hat{w}(y)$ and $w_6(y)$, respectively. 
Analogous results for the IB-corrected $\tau$ case
are shown in Figures \ref{fig1tau} and \ref{fig2tau}, respectively.
We see that the normalization and slope of the $\tau$ spectral
integrals are in excellent agreement with OPE expectations,
while both the normalization and slope are low in the EM case.
The same behavior is found for FESR's based on other non-negative,
monotonically decreasing weights, though for brevity we
have displayed only the $\hat{w}(y)$ and $w_6(y)$ results.
As noted above, this pattern is the one expected if it is the $\tau$ data
which are correct and the EM data which are wrong. (Note, however,
that an alternate $\tau$ spectral distribution which, like the
preliminary BELLE $\tau$ $\pi\pi$ distribution, is larger for some $s$ but 
smaller for other $s$ than is the corresponding ALEPH distribution,
is also capable of satisfying the OPE constraints; i.e., the fact that
the constraints are satisfied by the ALEPH data does not necessarily
imply that the ALEPH data set is correct.)

\begin{table}
\begin{center}
\caption{$\alpha_s(M_Z)$ from a fit to the EM and $\tau$ spectral 
integrals at $s_0\sim 4\ {\rm GeV}^2$ and $s_0\sim m_\tau^2$, respectively}
\end{center}
\begin{center}
\begin{tabular}{|ccc|}
\hline
$w(y)$&\ EM or $\tau$\qquad&$\alpha_s(M_Z)$\\
\hline
$\hat{w}$&EM&$0.1138^{+0.0030}_{-0.0035}$\\
$w_3$&EM&$0.1152^{+0.0019}_{-0.0021}$\\
$w_6$&EM&$0.1150^{+0.0022}_{-0.0026}$\\
\hline
$1-y$&$\tau$&$0.1212^{+0.0027}_{-0.0032}$\\
$w_3$&$\tau$&$0.1189^{+0.0018}_{-0.0021}$\\
$w_6$&$\tau$&$0.1195^{+0.0020}_{-0.0022}$\\
\hline
\end{tabular}
\end{center}
\label{table1}
\end{table}

\begin{table}
\begin{center}
\caption{The slopes with respect to $s_0$
of the EM OPE and spectral integrals. The 
first and second line OPE entries for each weight correspond
to the independent high-scale average and fitted values
of $\alpha_s(M_Z)$, respectively, as described in the text.}
\begin{tabular}{|lcc|}
\hline
$w(y)$&$S_{exp}$&$S_{OPE}$\\
\hline
$\hat{w}$&$.00872\pm .00026$&$.00943\pm .00008$\\
&&$.00934\pm .00008$\\
\hline
$w_6$&$.00762\pm .00017$&$.00811\pm .00009$\\
&&$.00805\pm .00009$\\
\hline
\end{tabular}
\end{center}
\label{table2}
\end{table}

\begin{table}
\begin{center}
\caption{Shifts in the effective EM spectral integrals 
associated with the $\pi\pi$ and $4\pi$ parts of the EM $\rightarrow$ 
$\tau$ spectral data replacement, for $s_0=2\, GeV^2 [m_\tau^2]$} 
\begin{tabular}{|lcc|}
\hline
$w(y)$&$\pi\pi$&$4\pi$\\
\hline
$\hat{w}$&$82\%\ [36\%]$&$18\%\ [64\%]$\\
$w_6$&$87\%$\ [45\%]&$13\%\ [55\%]$\\
\hline
\end{tabular}
\end{center}
\label{table3}
\end{table}

We now quantify the extent of the disagreement between
the EM spectral integrals and corresponding OPE expectations.
As a measure of the normalization disagreement, we fit 
$\alpha_s(M_Z)$ to the measured EM spectral integrals at
$s_0\sim 4\ {\rm GeV}^2$ (the highest scale for which it is 
still possible to construct the EM spectral function as a sum
over observed exclusive modes) and compare this result to the 
independent high-scale average noted above. A similar fit is
performed for the $\tau$ data, in this case at the maximum scale,
$s_0=m_\tau^2$, accessible in hadronic $\tau$ decay.
The results of these exercises are shown in Table~1.
The fitted values are seen to be in good agreement with
the high-scale average in the $\tau$ case,
but $\sim 2-2.5\, \sigma$ low in the EM case. Results in
agreement with those shown in the table are also obtained
for other non-negative, monotonically decreasing weights~\cite{otherweights}.

The results for the EM spectral integral and OPE slopes with respect to $s_0$ 
($S_{exp}$ and $S_{OPE}$, respectively) are shown 
in Table~2. In the OPE case two values are given
for each weight, that in the first row corresponding to the
independent high-scale average $\alpha_s(M_Z)=0.1198\pm 0.0020$,
that in the second row to the fitted value obtained above for the 
weight in question, and given already in Table~1. 
We see explicitly, as noted above, that the OPE slope is {\it very} 
insensitive to $\alpha_s(M_Z)$. As is evident from the table, no realistic
value for $\alpha_s(M_Z)$ will suffice to bring the OPE and spectral integral 
slopes into agreement; such agreement can only be obtained
through changes in the experimental spectral function. The 
slope discrepancies are at the $\sim 2.5\, \sigma$ level.

\begin{figure}
\unitlength1cm
\caption{OPE and $\tau$-modified ``EM'' spectral integrals for $w_6(y)$.
The solid circles (with error bars) represent the original EM
spectral integrals, the open circles the $\tau$-modified ``EM'' ones.}
\begin{minipage}[t]{8.0cm}
\begin{picture}(7.9,7.9)
\epsfig{figure=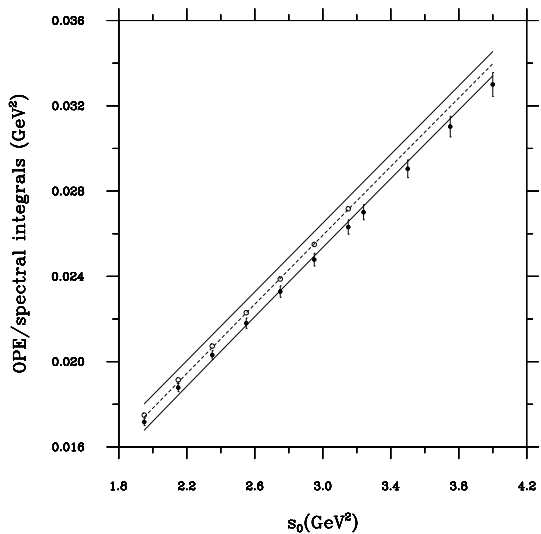,height=7.8cm,width=7.8cm}
\end{picture}
\end{minipage}
\label{figtaumodem}\end{figure}

To investigate the extent to which the source of the
normalization and slope disagreement in the EM case lies
in the $I=1$, as opposed to $I=0$, portion of the spectral function,
we replace the EM $\pi\pi$ and $4\pi$ data with the 
corresponding IB-corrected $\tau$ results, and rerun the
OPE/spectral integral comparison. It is found that both the
slope and normalization of the resulting $\tau$-modified ``EM'' spectral 
integrals are in excellent agreement with the OPE constraints. An 
illustration of this point, for the $w_6$ FESR,
is given in Figure~\ref{figtaumodem}.

\begin{table}
\begin{center}
\caption{Impact on the EM slope with respect to $s_0$ of the replacement 
of EM $4\pi$ data with the equivalent $\tau$ data, for $\hat{w}$ and $w_6$. 
The modified EM experimental slope ({\it exp}) is given in the first line,
the OPE slope ({\it OPE}) in the second.}
\begin{tabular}{|lcc|}
\hline
$w(y)$&$\alpha_s(M_Z)$&Slope \\
\hline
$\hat{w}$&$.1186$&$.00936\pm .00026$ (exp)\\
&&$.00940\pm .00008$ (OPE)\\
$w_6$&$.1176$&$.00795\pm .00017$ (exp)\\
&&$.00808\pm .00009$ (OPE)\\
\hline
\end{tabular}
\end{center}
\label{table4}
\end{table}

Finally, we consider the relative role of the $\pi\pi$ and $4\pi$ 
components of the $I=1$ $\tau$-EM spectral function difference
in producing the agreement between the OPE constraints
and the $\tau$-modified ``EM'' spectral integrals. Since 
$\left[ a_\mu\right]^{had,LO}$ is more strongly dominated by 
the $\pi\pi$ component of the spectral function than are the 
spectral integrals appearing on the LHS's of the FESR's employed 
in our analysis, this question is of relevance to determining the 
implications of our results for $\left[ a_\mu\right]^{had,LO}$.
 
The relative $\pi\pi$ and $4\pi$ contributions to the spectral integral shifts
caused by the $\rho_{EM}(s)\rightarrow\rho_\tau (s)$ replacement
are shown, for our two representative weights, in Table~3.
The results for the other $w_N$ are similar to those for $w_6$.
The $\pi\pi$ component is seen to dominate for $s_0\sim 2\ {\rm GeV}^2$
and remain important for $s_0\sim m_\tau^2$. 

Since the agreement between different experiments for the EM 
$\pi^+\pi^-\pi^0\pi^0$ cross-sections is, at present, not good~\cite{dehz}, 
it is of relevance to consider the impact of replacing only the $4\pi$
component of the EM spectral function with the corresponding
$\tau$ component. The impact of this change on the normalization
and slope of the EM spectral integrals is shown in Table~4.
We see that, were future results to bring the EM $4\pi$ cross-sections
into agreement with expectation based on $\tau$ decay data, the
difference between the slope and normalization of the EM spectral 
integrals and the OPE constraint values would be reduced to
the $\sim 1\, \sigma$ level. Such a change would also
reduce the discrepancy between the EM-data-based SM prediction
for $a_\mu$ and both the $\tau$-data-based SM prediction
and the E821 experimental result~\cite{ibfootnote}. The discrepancy between
the $\tau$ and EM versions of $\left[ a_\mu\right]^{had,LO}$
would also be reduced were the $\sim 4\times 10^{-10}$ reduction,
relative to the earlier $\tau$ average, seen in 
the preliminary BELLE result for the $\pi\pi$ contribution,
to remain present in the final version of the analysis. Such an improved
EM-$\tau$ consistency, according to the results of this study,
would almost certainly be accompanied by a reduction in the difference 
between the experimental $a_\mu$ value and the EM-based SM prediction.
Such a development would make even more important the role of
a reduced experimental uncertainty~\cite{newbnlproposal} 
in determining whether or not beyond-the-SM contributions 
have been detected in $a_\mu$.



\begin{thebibliography}{99}
\bibitem{otherweights}The results obtained for $\alpha_s(M_Z)$ 
using the $w_3,\, w_4$, and $w_5$ sum rules are, 
for the EM case, $0.1152^{+0.0019}_{-0.0021}$, $0.1154^{+0.0020}_{-0.0023}$,
and $0.1152^{+0.0022}_{-0.0024}$, respectively, and, for the $\tau$ case,
$0.1189^{+0.0018}_{-0.0021}$, $0.1193^{+0.0019}_{-0.0022}$,
and $0.1194^{+0.0020}_{-0.0022}$, respectively. All are in
excellent agreement with the corresponding result from the $w_6$ sum rule.
\end{thebibliography}
\end{document}